\documentstyle[aps,preprint]{revtex}
\draft

\begin{document}
\title
{An ansatz for the eigenstates in ${\cal PT}$-symmetric quantum mechanics}
\author{Zafar Ahmed}
\address{Nuclear Physics Division, Bhabha Atomic Research Centre,
Bombay 400 085\\
zahmed@barc.gov.in}
\date{\today}
\maketitle
\begin{abstract}
We suggest a general ansatz for the energy-eigenstates when a complex
one-dimensional ${\cal PT}$-symmetric potential possesses real discrete
spectrum. Several interesting features of ${\cal PT}$-symmetric quantum
mechanics have been brought out using this ansatz.
\end{abstract}
\vskip .3 in
The new option [1] that the non-Hermitian ${\cal PT}$-symmetric Hamiltonians may
too have real discrete spectrum has given rise to a lot of very interesting
investigations. Interestingly, in these developments exactly solvable models
[2-8] have not only shown the ways but also preceded the general proofs so much
that several general results are still unproved or unavailable.
\par The non-orthogonality of the new energy-eigenstate in the usual Hermitian sense
was first encountered in exactly solvable models (e.g., [4]). Then in several
independent [4,10-13] studies a new scalar product (${\cal PT}$-norm,
${\cal PT}$-orthogonality) was proposed. Even the indefiniteness of the new
norm which is well known now was first displayed only in an exactly solvable
model (e.g., [9])
\par So far it is not stated as to when expectation values of various operators
$x,f(x),p,p^2,H$ etc. under the new scalar product will be real. In the present
work we provide answer to this question by assuming an ansatz for the eigenstates
for a complex one-dimensional ${\cal PT}$-symmetric potential when it possesses real discrete
spectrum.
\par A general ${\cal PT}$-symmetric Hamiltonian can be written as
\begin{equation}
H={p^2_x \over 2m}+V_c(x)=-{d^2 \over dx^2}+V_0(x)+i\lambda V_1(x),
~~\hbar=1=2m, \lambda~ real,
\end{equation}
where real functions $V_0(x)$ and $V_1(x)$ are even and odd respectively.
We now propose a general ansatz for the $n^{th}$ eigenstate
of the Hamiltonian (1), as
\begin{equation}
\Psi_n(x)=\psi_{n,0}(x)+i\psi_{n,1}(x),~~s.t.~
\psi_{n,l}(-x)=(-)^{n+l}\psi_{n,l}(x),
\end{equation}
where $\psi_{n,l}(x)$ are real and essentially vanish at $x=\pm \infty~ or~ \pm L$.
In the above equation by noticing the novel parity scheme of $\psi_{0,1}(x)$, one
can check that the eigenfunctions of all the  exactly solvable models
complex one-dimensional ${\cal PT}$-symmetric potentials conform to this
interesting ansatz. In the following the indices (subscripts : $0,1,n$) will
be playing a very interesting role while appearing in the integrands of all
the integrals in the sequel. If sum of these indices is even the integral will
survive and will vanish otherwise.
\par
In one dimension, when the Hamiltonian and eigenfunctions are real we can
write $\cal{H^\ast}={\cal}H$, $\phi^\ast_n=\phi_n$ and $({\cal H}\phi_n)^\ast=
{\cal H}^\ast\phi^\ast_n={\cal H}\phi_n$ as the complex-conjugation operator
does not {\it transpose}.
In the same way, let us notice that $H^{PT}=H$ and $\Psi^{PT}_n=(-)^n \Psi_n$.
Let us write the eigenvalue equation for Eqs.(1) and (2), assuming that the
eigenvalues are complex $E_n=E^r_n+iE^i_n$,
\begin{equation}
H\Psi_n=(E^r_n+iE^i_n)\Psi_n.
\end{equation}
Noting that if $c$ is a constant then $(c)^{PT}=c^\ast$, we do ${\cal PT}$-operation
in Eq. (3) to get
\begin{equation}
H\Psi_n=(E^r_n-iE^i_n)\Psi_n.
\end{equation}
Subtracting Eq. (4) from (3), we find that $E^{i}_n=0$, showing that all
the eigenvalues of $H$ are real when the ${\cal PT}$-symmetry is exact:
the energy-eigenstate (2) is also the simultaneous eigenstate of the antiliear
operator ${\cal PT}$.
\par The expectation value of an operator $\hat O$ is usually defined as
\begin{equation}
<n|\hat O|n> ={1\over N_n}\int_{-\infty}^{\infty} \Psi^\ast _n(x)~
\hat O~ \Psi_n(x)~dx.
\end{equation}
When the boundary is at a finite distance, we will replace the limits
by $\pm L$. Let us also define the new expectation value as
\begin{equation}
(n|\hat O|n)={1 \over N^\prime_n}\int_{-\infty}^{\infty} \Psi_n(x)~
\hat O~ \Psi_n(x)~dx,
\end{equation}
where the complex conjugation has
been discreetly given up. In the following we shall be calculating both
$<n|\hat O|n>$
and $(n|\hat O|n)$ for various quantal operators to compare and contrast.
Since the ${\cal PT}$-symmetric potentials and their eigenstates corresponding
to real part of the spectrum recieve increasing attention in the most recent years,
the predictions as to whether the
expectation value or the just defined averages of various operators would be
real or imaginary, zero or non-zero and positive or negative would be valuable.
In the following we will see that the proposed ansatz and the arguments based
on parity of various entities would suffice to predict important results.
\par Let us calculate the norm, $N_n=<n|1|n>$
\begin{equation}
N_n=\int_{-\infty}^{\infty} \Psi^\ast_m(x) \Psi_n(x) ~dx=\int_{-\infty}^{\infty}
[\psi^2_{n,0}(x)+\psi^2_{n,1}(x)] ~dx.
\end{equation}
Let us also calculate a quantity $N^\prime_n=(n|1|n)$.
\begin{eqnarray}
N^\prime_n= \int_{-\infty}^{\infty} \Psi_n(x) \Psi_n(x) ~dx
&=&\int_{-\infty}^{\infty}
[\psi^2_{n,0}(x)-\psi^2_{n,1}(x)+2i\psi_{n,0}(x)~\psi_{n,1}(x)] ~dx
\nonumber \\
&=&\int_{-\infty}^{\infty} [\psi^2_{n,0}(x)-\psi^2_{n,1}(x)] ~dx.
\end{eqnarray}
In view of the interesting odd parity of the integrand (see Eq. (2)) the
the second integral vanishes.
This simply proves that ${\cal PT}$-norm ($N^\prime_n$) will be real but
indefinite (positive/negative) unlike the Hermitian norm ($N_n$) given in (7)
which is positive definite.
\par Now we calculate $<n|x^{2k}|n>, k=1,2...$ by noticing the parity of
various integrands in view of Eq.(2), we get
\begin{equation}
<n|x^{2k}|n>={1 \over N_n}\int_{-\infty}^{\infty} x^{2k} [\psi^2_{n,0}(x)
+\psi^2_{n,1}(x)]~dx
\end{equation}
which is real. Similarly, one can show that
\begin{equation}
<n|x^{2k+1}|n>=0
\end{equation}
On the other hand, using a similar analysis we find that
\begin{equation}
(n|x^{2k}|n)={1 \over N^\prime_n}\int_{-\infty}^{\infty} x^{2k}
[\psi^2_{n,0}(x)-\psi^2_{n,1}(x)]~dx
\end{equation}
is real.
Further,
\begin{equation}
(n|x^{2k+1}|n)={2i\over N^\prime_n}\int_{-\infty}^{\infty} x^{2k+1}
\psi_{n,0}(x)\psi_{n,1}(x)~dx,
\end{equation}
which is imaginary.
By noting that differentiation changes the parity of a definite parity function
and that $\psi_{n,l}(\pm \infty)=0,$ we find that
\begin{equation}
<n|p_x|n>={2\over N^\prime_n} \int_{-\infty}^{\infty} \psi^\prime_{n,0}(x)
\psi_{n,1}(x) ~dx,
\end{equation}
is real while $\psi^\prime_{n,0}(x)={d\psi_{n,0}(x) \over dx}$. But on the other hand, we find that
\begin{equation}
(n|p_x|n)=0
\end{equation}
As the double differentiation does not change the parity
of a definite parity function we use Eq.(3) to find that
\begin{equation}
<n|p^2_x|n>={1 \over N_n}\int_{-\infty}^{\infty} [E_n-V_0(x)][\psi^2_{n,0}(x)
+\psi^2_{n,1}(x)]~dx
\end{equation}
is real. Whereas
\begin{equation}
(n|p^2_x|n)={1\over N^\prime_n}\int_{-\infty}^{\infty} \{[E_n-V_0(x)]
[\psi^2_{n,0}(x)-\psi^2_{n,1}(x)] +2\lambda V_1(x)\psi_{n,0}(x)\psi_{n,1}(x)\}
~dx
\end{equation}
is also real. Next, we find the expectation value of $V_c(x)=V_0(x)+i\lambda
V_1(x)$, we have
\begin{equation}
<n|V_c(x)|n>={1\over N_n}\int_{-\infty}^{\infty} V_0(x)[\psi^2_{n,0}(x)
+\psi^2_{n,1}(x)]~dx
\end{equation}
and
\begin{equation}
(n|V_c(x)|n)={1 \over N^\prime_n}\int_{-\infty}^{\infty} \{V_0(x)
[\psi^2_{n,0}(x)-\psi^2_{n,1}(x)] -2\lambda V_1(x)\psi_{n,0}(x)\psi_{n,1}(x)\}
~dx.
\end{equation}
By adding Eq.(15) with Eq.(17) and Eq.(16) with Eq.(18), we prove that
\begin{equation}
<n|H|n>=E_n=(n|H|n) .
\end{equation}
This is a remarkable result which establishes the equality of expectation
values of the Hamiltonian under the old (5) and new (6) definitions.
We now take up the issue of the orthogonality of $\Psi_n(x)$.
For two states $m\ne n$ with distinct eigenvalues $E_m$ and $E_n$
(complex or real),
let us write the Schr{\"o}dinger equation as
%\begin{mathletters}
\begin{equation}
H\Psi_m(x)=E_m\Psi_m(x),
\end{equation}
\begin{equation}
H\Psi_n(x)=E_n\Psi_n(x).
\end{equation}
%\end{mathletters}
Let us left-multiply (20) by $\Psi_n(x)$  and (21) by $\Psi_m(x)$ and
subtract both the equations and integrate $w.r.t.x$ to have
\begin{eqnarray}
(E_m-E_n)\int_{-\infty}^{\infty} \Psi_m(x)\Psi_n(x)~dx \nonumber\\
=\int_{-\infty}^{\infty}[\Psi_n(x)H\Psi_m(x)-\Psi_m(x)H\Psi_n(x)]~dx
\nonumber \\
=-\int_{-\infty}^{\infty} \left \{\Psi_n(x){d^2 \over dx^2}\Psi_m(x)
-\Psi_m(x) {d^2 \over dx^2}\Psi_n(x)\right \}~dx  \nonumber \\
=\left [\Psi_n(x){d\Psi_m(x) \over dx}-\Psi_m(x){d\Psi_n(x) \over dx}
\right ]_{-\infty}^{\infty}~=~0.
\end{eqnarray}
The last term vanishes due to the boundary condition i.e.,
$\Psi_j(\pm\infty)=0$.
And so we have proved the orthogonality of two eigenstates corresponding
to two distinct
eigenvalues for a ${\cal PT}$-symmetric Hamiltonian in general.
Thus, the orthogonality condition for the eigenstates (2) of an arbitrary
${\cal PT}$-invariant potential can now be stated as
\begin{equation}
(E_m-E_n)\int_{-\infty}^{\infty} \Psi_m (x)~\Psi_n (x)~dx=0,~m \ne n,
\end{equation}
notice the absence of complex conjugation.
Using Eqs. (20) and (21) and bearing in mind that the complex-conjugation
does not {\it transpose}, we can derive
\begin{eqnarray}
(E_n-E_m)\int_{-\infty}^{\infty}\Psi^\ast_m(x) \Psi_n(x)~dx
=2i\int_{-\infty}^{\infty} \Psi^\ast_m(x) \lambda V_1(x) \Psi_n(x)~dx
\nonumber \\
=2\int_{-\infty}^{\infty}\lambda V_1(x)[\psi_{m,1}(x) \psi_{n,0}(x)
-\psi_{m,0}(x)\psi_{n,1}(x)]~dx, if~ m+n=even, \nonumber \\
=2i\int_{-\infty}^{\infty} \lambda V_1(x)[\psi_{m,0}(x)
\psi_{n,0}(x)+\psi_{m,1}(x) \psi_{n,1}(x)]~dx, if~ m+n=odd.
\end{eqnarray}
It is helpful to note that whenever $\psi_{p,q}(x)$ appears in an integral
and if sum of all the indices of the integrand is even the integral survives
and the integral vanishes otherwise.
The Eq.(24) demonstrates that $\Psi_m(x)$ and $\Psi_n(x)$ are not orthogonal
in the conventional way.
For a better insight, let us see the orthogonality in another way by expressing
$<m|n>$ and $(m,n)$ as
\begin{eqnarray}
<m|n>=\int_{-\infty}^{\infty} \Psi^\ast_m(x)~\Psi_n(x)~dx,
=\int_{-\infty}^{\infty}\{[\psi_{m,0}(x)\psi_{n,0}(x)+\psi_{m,1}(x)\psi_{n,1}(x)] \nonumber              \\
+i[\psi_{m,0}(x)\psi_{n,1}(x) -\psi_{m,1}(x)\psi_{n,0}(x)]\}~dx,
\end{eqnarray}
\begin{eqnarray}
(m|n)=\int_{-\infty}^{\infty} \Psi_m(x)~\Psi_n(x)~dx,
=\int_{-\infty}^{\infty} \{[\psi_{m,0}(x)\psi_{n,0}(x)-\psi_{m,1}(x)\psi_{n,1}(x)]
\nonumber \\
+i[\psi_{m,0}(x)\psi_{n,1}(x)+\psi_{m,1}(x)\psi_{n,0}(x)]\}~dx.
\end{eqnarray}
A comparison of Eq. (26) with Eq. (23) leads to two interesting
additional properties of $\psi_{p,q}(x)$, viz.,
%\begin{mathletters}
\begin{equation}
\int_{-\infty}^{\infty} \psi_{m,0}(x)\psi_{n,0}(x)~dx=\int_{-\infty}^{\infty}
\psi_{m,1}(x)\psi_{n,1}(x)~dx, ~m \ne n,
\end{equation}
and
\begin{equation}
\int_{-\infty}^{\infty} \psi_{m,0}(x)\psi_{n,1}(x)~dx=-\int_{-\infty}^{\infty}
\psi_{m,1}(x)\psi_{n,0}(x)~dx, ~m \ne n.
\end{equation}
if the ansatz (2) were to represent the eigenstate of a ${\cal PT}$-symmetric
Hamiltonian with real discrete spectrum.
More importantly, notice once again that the usual orthogonality condition
using the complex-conjugation would not hold as Eq. (26) does not vanish in
view of properties Eqs.(27) and (28).
It is crucial to note that the right hand sides of Eqs. (24)
and (25) would vanish due to different set of conditions. A common
but trivial condition would of course be when $\lambda=0$ implying that
the potential is real. It is also imprtant to notice that
\begin{equation}
\int_{-\infty}^{\infty} \psi^2_{n,0}(x)~dx \ne \int_{-\infty}^{\infty}
\psi^2_{n,1}(x)~dx.
\end{equation}
from Eq. (27). This means that ${\cal PT}$-norm (8) in case of real discrete
spectrum does not vanish.
%\end{mathletters}
\par
So far we have worked with the proposed ansatz (2) which represnts only the
scenario when the ${\cal PT}$-symmetry is exact: the energy-eigenstates are also the
eigenstaes of the anti-linear operator ${\cal PT}$ and the discrete energy-eigenvalues
are real. We now consider the general case when the ${\cal PT}$-symmetry could be broken
or unbroken. By operating with ${\cal PT}$ in both the sides and remembering that
$(c)^{PT}=c^\ast$ and the fact that complex-conjugation does not {\it transpose}, using Eqs. (20,21)
we find that
\begin{eqnarray}
(E^\ast_m-E_n)\int_{-\infty}^{\infty} \Psi^{PT}_m(x)\Psi_n(x)~dx \nonumber \\
=\int_{-\infty}^{\infty}[\Psi_n(x)H\Psi^{PT}_m(x)-\Psi^{PT}_m(x)H\Psi_n(x)] ~dx
\nonumber \\
=-\int_{-\infty}^{\infty} \left \{\Psi_n(x){d^2 \over dx^2}\Psi^{PT}_m(x)
-\Psi^{PT}_m(x) {d^2 \over dx^2}\Psi_n(x)\right \}~dx,  \nonumber \\
=\left [\Psi_n(x){d\Psi^{PT}_m(x) \over dx}-\Psi^{PT}_m(x){d\Psi_n(x) \over
dx} \right ]_{-\infty}^{\infty}~=~0.
\end{eqnarray}
It is could be instructive to notice that the just derived orthogonality
condition i.e.,
\begin{equation}
(E^\ast_m-E_n) \int_{-\infty}^{\infty} \Psi^{PT}_m(x)~\Psi_n(x)~dx~=~0.~
\end{equation}
is more general as it degenerates to (24) when the ${\cal PT}$-symmetry is
unbroken i.e., $\Psi^{PT}(x)= (-)^n\Psi(x)$, and the eigenvalues are real,
$(E^\ast_m=E_m)$. When $m=n$ and $\Psi_m(x)$ are no more eigenstates of
${\cal PT}$, the ${\cal PT}$-symmetry is broken and eigenvalues are complex conjugate
pairs and consequently according to Eq. (31) the ${\cal PT}$-norm vanishes.
Such states are also termed as self-orthogonal states [14].
Following the suggestions in Refs. [4,10-13], we could have also defined the
expectation value $(n|\hat O|n)$ (8) alternatively as
\begin{equation}
(n|\hat O|n)={1 \over N^\prime_n} \int_{-\infty}^{\infty} \Psi^{PT}_n(x) \hat O \Psi_n(x)~dx,
N^\prime_n=\int_{-\infty}^{\infty} \Psi^{PT}_n(x) \Psi_n(x)~dx.
\end{equation}
Since $\Psi^{PT}_n(x)=(-)^n \Psi_n(x)$ in the case of the real discrete spectrum,
the above results remain unchanged. An interesting account of the definition of
the expectation (6) in the light of bi-orthogonality can be found in a very recent
work [15].
\par The results proved in Eqs. (9-12,14,16,18) can be summarized by stating
that in ${\cal PT}$-symmetric quantum mechanics only ${\cal PT}$-symmetric
operators can have real
expectation values. This means the expectation values of $ix$ and $e^{ix}$ will
be real. This is akin to the conventional Hermitian  quantum mechanics wherein
only Hermitian operators have real eigenvalues. The result in Eq. (19) is
interestingly surprising which states that expectation value of the complex
${\cal PT}$-symmetric Hamiltonian remains equal to the energy eigenvalue
irrespective of
the definitions of the expectation value given in Eq. (5) and (6). This means the
variational method [10] can work for the new Hamiltonians under both old and new
definitions. We remark that these results have not appeared before.
Eq.(8) proves the indefiniteness of the ${\cal PT}$-norm. Importantly,
Eq. (29) proves that in the case of unbroken ${\cal PT}$-symmetry (real discrete
spectrm) the new norm does not vanish.
\par We would like to re-emphasize that all these results could be proved
merely by the parity scheme of the suggested ansatz for the energy-eigenfunction
in Eq. (2). It may be verified that the energy-eigenfunction of all the exactly
solvable one-dimensional models of the complex ${\cal PT}$-symmetric potentials
essentially conform to the suggested ansatz (2).
\section *{References : }
\begin{enumerate}
\item C.M.Bender and S. Boethcher, Phys. Rev. Lett. {\bf 80} (1998) 5243.
\item M. Znojil, Phys. Lett. A {\bf 259} (1999) 220.
\item G. Levai and M. Znojil, J.Phys. A: Gen. Math. {\bf 33} (2000) 7165.
\item Z. Ahmed, Phys. Lett. A {\bf 282}(2001) 343; {\bf 287} (2001) 295.
\item H.F. Jones, Phys. Lett. A {\bf 262} (1999) 242.
\item Z. Ahmed, Phys. Lett. A {\bf 286} (2001) 231.
\item Jia C-S, Zeng X-L, Sun L-T, Phys. Lett. A {\bf 300} (2002) 78.
\item A. Sinha, G. Levai, P. Roy, J. Phys. A: Math. Gen. {\bf 36} (2003) 7611.
\item B. Bagchi, C. Quesne, M. Znojil, Mod. pgys. Lett. A {\bf 16} (2001) 2047;
G. Levai, F. Cannata and A. Ventura, phys. Lett. A {\bf 300} (2002) 271.
\item C.M. Bender, F. Cooper, P.N. Meisinger, V.M. Savage, Phys. Lett. A 259
(1999) 224.
\item M. Znojil, math-ph/0104012.
\item G.S. Japaridze, J. Phys. A: Math. Gen. {\bf 35} (2003) 1709.
\item C.M. Bender, D.C. Brody and H.F. Jones, Phys. Rev. Lett. {\bf 89} (2002) 270401.
\item E. Narevicius, P. Serra and N. Moiseyev, Eur. Phys. Lett. {\bf 62} (2003) 789.
\item A.V. Sokolov, A.A. Andrianov, F. Cannata, quant-ph/0602207.
\end{enumerate}
\end{document}